\documentclass[twocolumn,showpacs,preprintnumbers,superscriptaddress,prl]{revtex4}
\usepackage{graphicx}
\newcommand{\la}{\langle}
\newcommand{\ra}{\rangle}
\newcommand{\btem}{\bibitem}

\begin{document} 
\title{Scalar Mesons  in Lattice QCD}

%
\author{Teiji Kunihiro}
\affiliation{YITP, Kyoto University, Kyoto 606-8502, Japan}
\author{Shin Muroya}
\affiliation{Tokuyama Women's College, Shunan 745-8511, Japan}
\author{Atsushi Nakamura}
\affiliation{RIISE, Hiroshima University, Higashi-Hiroshima 
739-8521,  Japan }
\author{Chiho Nonaka}
\affiliation{Department of Physics, Duke University, Durham, 
NC 27708-0305, USA}
\author{Motoo Sekiguchi}
\affiliation{Faculty of Engineering, Kokushikan University,
Tokyo 154-8515, Japan}
\author{Hiroaki Wada}
\affiliation{Laboratory of Physics, College of Science and
Technology, Nihon University, Chiba 274-8501, Japan}
       \collaboration{SCALAR collaboration} \noaffiliation

\begin{abstract}
We explore whether ``f$_{0}$(600) or $\sigma$" exists as a pole in QCD,
by a full lattice QCD simulation on the $8^{3} \times 16$ lattice using
the plaquette action and Wilson fermions.
It is  shown that a low-mass  $\sigma$ pole exists 
owing to the disconnected diagram of the meson 
propagator. 
A  discussion is given on the physical content of the $\sigma$.
\end{abstract}


\maketitle


Recently there has been renewed and growing interest in scalar mesons 
 \cite{KEK,close-tornqvist}.
One of the most noteworthy developments in this field is 
the re-identification of the $\sigma(600)$ after some twenty
 years.
 The results of re-analyses of the $\pi$-$\pi$ scattering phase shift 
 of the scattering ($S$) matrix strongly suggest the existence 
 of the pole of the $\sigma$  meson with $I=0$ and $J^{PC}=0^{++}$
 in the $s$ channel as well as the $\rho$ pole in the $t$ channel.
In  confirming the $\sigma$ pole, it was essential  
to respect chiral symmetry,  analyticity, unitarity 
and crossing symmetry in constructing the $S$
matrix \cite{crossing,OOR,ChiralPurt}.
The significant  contributions of the $\sigma$ pole
were also identified in the decay processes
of heavy particles such 
as D$\to \pi \pi \pi$ and $\Upsilon(3S) \to \Upsilon \pi \pi$ 
\cite{KEK,decay,Ishida,bugg}.
The meson called  ``$f_0$(600) or $\sigma$"  appeared 
in the 2002 edition of PDG \cite{PDG02}.
It should be noted that a recent analysis \cite{argand}  clearly 
shows phase motion consistent with resonance behavior.

As stressed above, the significance of the $\sigma$ pole 
is intimately related to chiral symmetry in QCD; 
the $\sigma$ may be identified as the chiral partner of 
the $\pi$ in the linear representation of chiral
SU$(2)\otimes$SU$(2)$ symmetry  \cite{GL60}. 
Here, it is noteworthy that the $\kappa$ meson with $I=1/2$
is reported to exist with a mass $m_{\kappa}$ of about 
800 MeV \cite{E791,BES,bugg}; 
the $\kappa$ is supposed to constitute
the nonet scalar states of chiral SU$(3)\otimes$SU$(3)$ symmetry
together with the $\sigma$.
See `Meson Particle Listings $f_0(600)$' 
in Ref. \cite{PDG02} (pp.010001-450 $\sim$ 453), 
and references therein.

Even with the confirmed existence of the low lying scalar mesons, 
there are long standing controversies on their nature
 \cite{close-tornqvist}:
 Are they usual $q\bar{q}$ mesons,
four quark states  like $qq\bar{q}\bar{q}$ \cite{jaffe},
 or $\pi\pi$ ($K\bar{K}$)
molecules, rather than  pre-existing resonances? 
How large is the mixing among them or with glueballs \cite{narison}? 
Although not mentioned in 
\cite{close-tornqvist}, 
the $\sigma$  may be a {\em collective} $q\bar{q}$ state described
as a superposition of  many {\em atomic} $q\bar{q}$ 
states \cite{NJL,HK85}:
In fact, the dynamical models of
chiral symmetry breaking, such as the Nambu-Jona-Lasinio model
\cite{NJL}
and the Schwinger-Dyson approaches  \cite{ES84},
describe  the $\sigma$ 
as well as the $\pi$ as collective states.

Confronting these problems with the $\sigma$,
it would be  intriguing to explore whether 
QCD accommodates the $\sigma$ and other
possible low-lying scalar mesons, as well as 
to examine  their nature by a first-principles calculation of QCD.

DeTar and Kogut  \cite{DeTar} were the first who  measured the  
screening mass of the $\sigma$ meson in lattice QCD, 
although in  the quenched approximation.
In the same approximation, Alford and Jaffe explored 
possible light scalar mesons  described as
$q^2\bar{q}^2$ states  \cite{Alford}, while
the masses of the $q\bar{q}$ states and their  mixing with 
glueball states in the scalar channel
were investigated by Lee and Weingarten \cite{Lee}.  
It was pointed out  \cite{Lat01} that  the so-called 
disconnected diagram, which is not properly treated in the 
quenched approximation,
gives almost the same amount of  contribution  to the propagator
as the connected diagram does; this 
indicates that using the dynamical quarks is essential 
to study the $\sigma$ on a lattice.

A simulation with dynamical quarks was carried out by 
McNeile and Michael \cite{McNeile}, who computed the masses of
the iso-singlet scalar states given by a superposition of
$q\bar{q}$ and glueball states. 
In their simulation, the $\sigma$ meson was found to be lighter than the $\pi$;
hence the relevance of the simulation in reality is obscure.

Since the existence and the nature of the $\sigma$ may be
related to chiral symmetry and its dynamical breaking,
it is desirable to examine the behavior of the chiral limit of the
calculated quantities for a proper description of  the
$\sigma$ meson. 
Of course, it is preferable to adopt lattice fermions which 
have better chiral properties.
Using the domain wall fermions,
RBC (Riken-Brookhaven-Columbia) collaboration
examined the $\sigma$ propagator on the lattice,
unfortunately, however, in the quenched approximation
and with an approximate estimate of
the disconnected diagram based on 
the chiral perturbation theory  \cite{Sasa}.

In this letter, we present a lattice calculation of the 
scalar particles using full QCD with the inclusion of 
the disconnected diagrams.
We employ the most standard lattice QCD numerical techniques, 
and show that there exists a $\sigma$ pole,
whose mass is similar to that of  the $\rho$ meson for the
quark masses used for the simulation but tends to become
as low as $2m_{\pi}$ in the chiral limit.
We shall show that the incorporation
of the disconnected diagram is responsible 
for realizing such a low mass of the $\sigma$ meson. 
We shall briefly discuss the physical content of the $\sigma$ meson.
We shall also mention that our simulation suggests the existence of 
the $\kappa$ meson, another member of the nonet of the scalar
mesons.

We adopt the following interpolation operator
for creating the $\sigma$ meson with $I=0$ and 
$J^{PC}=0^{++}$,
\begin{equation}
\hat{\sigma}(x) \equiv 
\sum_{c=1}^3\sum_{\alpha=1}^4 
\frac{\bar{u}_\alpha^c(x)u_\alpha^c(x)+\bar{d}_\alpha^c(x)d_\alpha^c(x)}
{\sqrt{2}} ,
\label{eq:sigma_operator}
\end{equation}
where $u$ ($d$) denotes the up-quark (down-quark) operator 
with  $c$ and $\alpha$ 
being the color and Dirac-spinor indices, respectively.
Then the $\sigma$ meson propagator reads
\begin{eqnarray}
G(y,x) &=&  
\la \mbox{T} \hat{\sigma}(y) \hat{\sigma}(x)^\dagger \ra 
\nonumber \\
&=& \frac{1}{Z} \int DUD\bar{u}DuD\bar{d}Dd 
\nonumber \\
& & \times 
\ \hat{\sigma}(y) \hat{\sigma}(x)^\dagger \ e^{-S_G-\bar{u}Wu-\bar{d}Wd}. 
\label{eq:propa_def}
\end{eqnarray}
Here $W^{-1}$'s are the $u$- and $d$-quark propagators, 
$U$'s the link variables of the gluon, 
and $S_G$ the gauge action. 
By integration over 
the quark fields,
the $\sigma$ meson propagator is reduced to 
\begin{eqnarray}
G(y,x) 
%
&=& - \la \mbox{Tr} W^{-1}(x,y) W^{-1}(y,x) \ra 
\nonumber \\
& & + 2 \la ( \sigma(y) - \la \sigma(y) \ra )
( \sigma(x) - \la \sigma(x) \ra ) \ra ,
\label{eq:propa}
\end{eqnarray}
where 
\begin{equation}
\sigma(x) \equiv \mbox{Tr} W^{-1}(x,x) ,
\nonumber
\end{equation}
with ``Tr'' 
being the trace operation over the color and Dirac indices. 
Here we set the $u$ and $d$ quark propagators to be  the same. 
Equation (\ref{eq:propa}) shows that 
the $\sigma$ propagator consists of two terms:
one corresponds to a connected diagram
and the other to a disconnected diagram.
Since the vacuum expectation 
value 
$\la \sigma(x) \ra$ does not vanish,
it should be subtracted from the $\sigma$ operator.
We employ Wilson fermions and the plaquette gauge action. 
We perform a full QCD simulation in which
the disconnected diagrams are included.

As for the simulation parameters,
we first note that CP-PACS performed a full-QCD calculation
of the light meson spectroscopy with great success \cite{CPPACS}. 
Therefore, we use the same values of the simulation parameters
as those used by CP-PACS, 
{\it i.e.},  $\beta = 4.8$ and the hopping parameter, $\kappa = 0.1846$, 
$0.1874$ and $0.1891$, except for the lattice size; 
our lattice size is  $8^3\times16$.
We shall use the point source and sink,
which together with the smaller lattice size,
leads to larger masses due to a mixture of higher mass states.  
In other words, the masses to be obtained in our simulation
should be considered as upper limits. 
We show in Table 1 the value of $m_\pi/m_\rho$ together with 
that of CP-PACS.
They are consistent within the error bars. 
\begin{table}[htb]
\begin{center}
Table 1:  Summary of the results. 
\begin{tabular}{c|c|c|c}
\hline
\hline
$\kappa$& 0.1846 & 0.1874 &0.1891 \\ \hline
statistics $^{1)}$         & 1110        &  860       & 730  \\ \hline
$m_{\pi}/m_{\rho}$ $^{2)}$ & 0.8291(12)  &  0.7715(17) & 0.7026(32) \\ \hline
$m_{\pi}/m_{\rho}$ $^{3)}$ & 0.825(2)    &  0.757(2)   & 0.693(3) \\ \hline
$m_{\sigma}/m_{\rho}$ $^{3)}$ & 1.6(1)     & 1.34(8)     & 1.11(6) \\ \hline
$m_{\rm connect}/m_{\rho}$ $^{3)}$ & 2.40(2)     & 2.44(3)     & 2.48(4) \\ \hline
\end{tabular} \\
$^{1)}$Number of configurations separated by 10 trajectories to each other. 
$^{2)}$CP-PACS. 
$^{3)}$Our result.
$m_{\rm connect}$ is the $\sigma$ mass estimated only from the connected
part.
\end{center}
\end{table}
\vspace{-2mm}
We generate the gauge configurations in full QCD using 
the hybrid Monte Carlo (HMC) algorithm.
The first 1500 trajectories are updated in quenched QCD,
then we switch to a simulation with
the dynamical fermion. 
The next 2000 trajectories 
of HMC are discarded for thermalization and 
the $\sigma$, $\pi$ and $\rho$ propagators are calculated 
every ten trajectories.

\begin{center}
\begin{figure}[htb]
\rotatebox[origin=c]{-90}{
\includegraphics[width=.6 \linewidth]{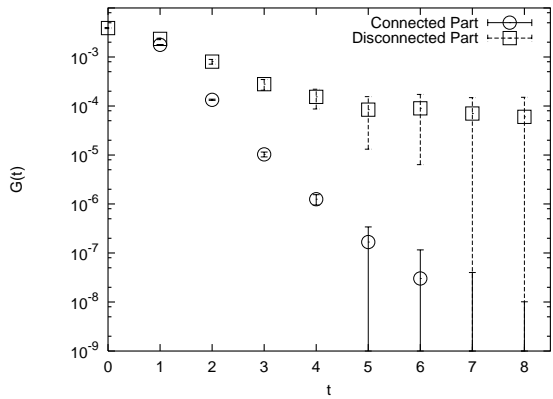} 
}
\vspace{-1.0cm}
\caption{
Propagators of the connected and disconnected diagrams of 
the $\sigma$ for $\kappa=0.1874$.}
\label{fig:ConDis_k1874}
\end{figure}
\end{center}

It is not an easy task 
to evaluate the disconnected part of the propagator,
since one must calculate $\mbox{Tr}W^{-1}(x,x)$ for all lattice sites $x$.
We used the $Z_2$ noise method to calculate the disconnected diagrams,
$\sigma(x)$, 
and the subtraction terms of the vacuum $\la \sigma \ra$.
Each of these terms is the order of ten, and 
$\la (\sigma-\la\sigma\ra)(\sigma-\la\sigma\ra)\ra$ becomes less than
$10^{-4}$, as shown in Fig.\ref{fig:ConDis_k1874}.
Therefore, in order to obtain the signals  correctly 
as the difference between these terms, 
numerical simulations with a high precision and careful analyses are 
necessary.
One thousand random $Z_2$ numbers are generated for each configuration:
We refer to our previous report  \cite{Lat01} for 
the relationship between the amount of $Z_2$ noise and the achieved
accuracy.
Gauge configurations are created by HMC in a vector supercomputer,
while most of the disconnected propagator is calculated
in a parallel machine. 

\begin{center}
\begin{figure}[htb]
\rotatebox[origin=c]{-90}{
\includegraphics[width=.6 \linewidth]{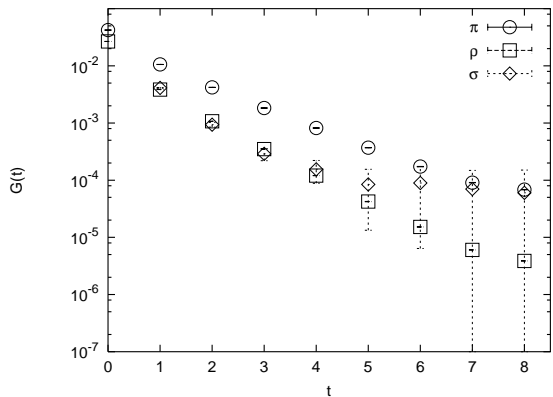} 
}
\vspace{-1.0cm}
\caption{Propagators of the $\pi$, $\rho$ and  $\sigma$ for 
$\kappa=0.1874$.}
\label{fig:PiRhoSig_k1874}
\end{figure}
\end{center}

The propagators of the $\pi$, $\rho$ and $\sigma$ for $\kappa=0.1874$ 
are shown in Fig.\ref{fig:PiRhoSig_k1874}.
From our results, we 
estimate the critical value of the hopping parameter 
and the lattice space as $\kappa_c = 0.195(3)$ and  $a = 0.207(9)$ fm,
respectively, which are to be compared with 
the CP-PACS values, 
$\kappa_c = 0.19286(14)$ and $a = 0.197(2)$ fm. 
The mass ratios of $m_\pi/m_\rho$ and $m_\sigma/m_\rho$ 
together with $m_{\rm connect}/m_\rho$
are summarized in Table 1, where $m_{\rm connect}$ is the mass
evaluated only from the connected diagram. 
We 
have extracted the  $\pi$ and $\rho$ masses using two-pole formula from
their propagators.  
The small errors of the $\pi$ and $\rho$ propagators indicate 
the high precision of our simulation.  
Our result for the mass ratios 
$m_{\pi}/m_{\rho}$ in Table 1 is in  a good agreement
with that of  CP-PACS, 
with unavoidable small differences
owing to the different lattice sizes.
To extract the $\sigma$ meson mass, 
we have used the one-pole ansatz, since its
propagator has large error bars at large $t \sim 6$
in spite of our high statistics;
it implies that the obtained value gives
an upper limit of the scalar  meson mass.
Figure \ref{fig:PiRhoSig_k1874} and Table 1 show 
that the mass of the $\sigma$
is of the same order as the mass of the $\rho$;
we remark that the $2\pi$ threshold is higher than the 
$\sigma$ in our simulation. 
We have checked that 
the effective mass method
gives essentially the same result as that given above.

Figure \ref{fig:ConDis_k1874} shows the individual contributions 
of the connected and disconnected parts of the $\sigma$ propagator,
which  reaches a plateau at $t \ge 5$, 
since the precision of our calculation is limited to
{\it O}($G(t))\sim10^{-4}$.
The connected part shows clear signals of a rapid damping 
with small error bars.  
We note that the connected part of 
the $\sigma$ propagator can be regarded as the
one  of the $a_{0}$ meson 
provided that the difference between the $u$-quark
and the $d$-quark is neglected. 
Therefore, the rapid damping of the connected part of the 
$\sigma$ propagator
is in accordance with the fact that the $a_{0}$ is 
heavier than the $\rho$.
Figure \ref{fig:ConDis_k1874} shows that
the disconnected part dominates the $\sigma$ propagator. 
By comparing Fig.\ref{fig:PiRhoSig_k1874} with Fig.\ref{fig:ConDis_k1874}, 
we see that the
$\sigma$ as a light meson results from the  disconnected 
part of the $\sigma$ propagator with  the background vacuum 
condensate subtracted, which are odd with the 
constituent quark model. 
In the naive sense, the connected quark diagram corresponds to
the constituent quarks in the SU(3) non-relativistic quark model 
where the OZI rule is 
satisfied.
This may give a clue to clarify the physical 
content of the $\sigma$ meson, as will be discussed later.
The mass of the connected $\sigma$ (the $a_{0}$) shown in Table 1
is almost 2.5 times  of the $\rho$ mass
and exhibits only a weak dependence on the hopping parameter,
suggesting an irrelevance of chiral symmetry to the
$a_0$ meson.  

\begin{center}
\begin{figure}[htb]
\rotatebox[origin=c]{-90}{
\includegraphics[width=.6 \linewidth]{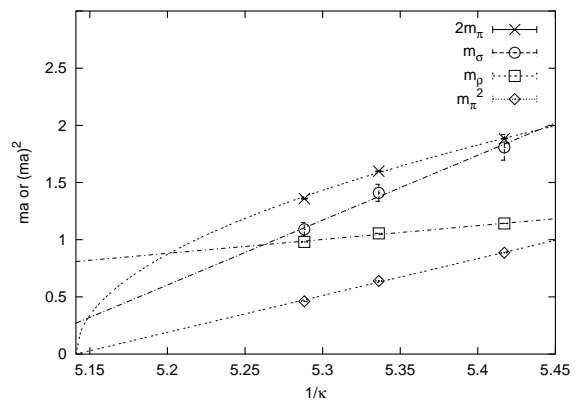}
}
\vspace{-1.0cm}
\caption{
$m_\pi^2$, $m_\rho$, $m_\sigma$
and $2m_\pi$ in the
lattice unit as a function of the inverse hopping parameter. 
The chiral limit is given by $\kappa_c=0.195(3)$.
}
\label{fig:Extrapolate}
\end{figure}
\end{center}

In Fig.\ref{fig:Extrapolate},  we display $m_\pi^2$, $m_\rho$, $m_\sigma$
and $2m_\pi$ in lattice units as a function of the inverse 
hopping parameter. 
As the chiral limit is approached, 
the $\sigma$ meson 
mass obtained from  the full $\sigma$ propagator decreases
and eventually becomes smaller than the $\rho$ meson mass 
 in the chiral limit. 

We have also calculated  the $\kappa$ meson using the same 
configurations. We take the same hopping parameter values 
for $u$ and $d$ quarks
and use the following three values for 
the $s$ quark hopping parameter, 
$\kappa_s$ = 0.1835, 0.1840 and 0.1845. 
The ratios $m_\kappa/m_{K^*}$ and $m_K/m_{K^*}$ at the chiral limit are 
listed in Table 2.
One may notice that 
all the three values of  $\kappa_s$
fairly reproduce the physical
mass ratio, $m_{K}/m_{K^*}$
in the  chiral limit of the $u$- and $d$-quarks. 
A more notable point is that 
the resultant $\kappa$ mass hardly changes with the variation
of $\kappa_s$ and is almost twice as heavy as the $K^{*}$; {\it i.e.},
the $\kappa$ is  not degenerated with the $\sigma$,  
although they should belong to the same nonet.
The origin to lift the octet-singlet degeneracy 
may be attributed the following facts;
because of the strangeness content, the $\kappa$ propagator
is solely composed of a connected diagram and 
contains no disconnected part, 
the latter of  which
was the origin of the light mass of the $\sigma$. 
In fact,
the $\kappa$ mass listed in Table 2 is almost the same as the $a_0$ mass
estimated using the connected part of the $\sigma$ with the corresponding
hopping parameters. 
Thus we can 
understand why the $\sigma$ is light while the $\kappa$ is heavy,
consistently.

\begin{table}[htb]
Table 2:  The mass ratios $m_\kappa/m_{K^*}$ versus $m_K/m_{K^*}$ 
at $\kappa_c = 0.1945(29)$. 
The $s$ quark hopping parameters were taken to be 
$0.1835$, $0.1840$ and $0.1845$. \\
\begin{tabular}{c|c|c|c} 
\hline 
\hline 
$s$ quark hopping parameter & 0.1835 & 0.1840 & 0.1845 \\ \hline
$m_K/m_{K^*}$       & 0.639(6)  &  0.631(6)  &  0.623(6)  \\ \hline
$m_{\kappa}/m_{K^*}$   & 2.039(43)  &  2.037(43)  &  2.044(44)  \\ \hline
\end{tabular} \\
\end{table}
\vspace{-2mm}


The dominance of the disconnected diagram in the $\sigma$ meson propagator
in contrast to the octet scalar mesons
suggests possible physical contents of the light $\sigma$: 
The disconnected diagram includes the process
$q\bar{q}\quad \leftrightarrow  G$ where  
$G$ denotes a glueball states.
This might  suggest an importance of the 
glueball mixing in the light $\sigma$ if  the glueball states were
not heavy.
Instead, we notice that such a fluctuation process of 
the $q\bar{q}$ states with the heavy $G$ giving an 
effective vertex for $q\bar{q} \rightarrow q\bar{q}$ 
 may form a collective  state mentioned before. 
One may also  notice that the disconnected diagram
contains  four quark states 
composed of diquark  and anti-diquark states, {\it i.e.},
$qq$-$\bar{q}\bar{q}$.
Clearly more works are needed to elucidate the physical
content of the scalar mesons.


We have reported the first study of the scalar mesons 
based on the full QCD lattice  simulation with dynamical fermions,
including analysis of the disconnected diagram effects.
We have used the most standard lattice QCD techniques, which have worked
well for the established mesons and baryons, and
clarified the accuracy and statistics required to obtain signals
in the scalar channel.

Our results indicate the existence of a light isoscalar
$J^{PC}=0^{++}$ scalar meson, {\it i.e.}, the $\sigma$ meson 
with a mass of almost the same order as that 
of  the $\rho$ meson (see Table 1). We have also 
shown that the $\kappa$ meson belonging to the
flavor octet is 
almost twice as heavy as the $K^{*}$.

Of course, the results reported here are in far from final 
quantitative stage because our quark masses are much heavier than 
those in the real world and our scalar mesons cannot decay on our lattice.
Nevertheless, it is now clear that the $\sigma$ meson as
well as other scalar mesons can be studied
in the lattice QCD.
We hope that the present
work prompts other lattice studies which will give
a deeper and more quantitative understanding of
the scalar mesons, especially the $\sigma.$

\noindent
{\bf Acknowledgment}
This work is supported by Grants-in-Aid for Scientific Research by
Monbu-Kagaku-sho (No.\ 11440080, No.\ 12554008, No.\ 12640263 and
No.\ 14540263) and ERI of Tokuyama Univ and DOE grants DE-FG02-96ER40495.
Simulations were performed on SR8000 at IMC, Hiroshima
Univ., SX5 at RCNP, Osaka Univ., and SR8000 at KEK.

\vspace{-4mm}

\end{document}